\documentclass{article}

\usepackage{graphicx}

\begin{document}

\def\be{\begin{equation}}
\def\ee#1{\label{#1}\end{equation}}

\title{Viscous cosmological models and accelerated Universes} 
\author{G. M. Kremer\thanks{kremer@fisica.ufpr.br}  $\,$ and 
F. P. Devecchi\thanks{devecchi@fisica.ufpr.br}\\
Departamento de F\'\i sica, Universidade Federal do Paran\'a\\
Caixa Postal 19044, 81531-990, Curitiba, Brazil
}
\maketitle
\date{}
\begin{abstract}
It is shown that a present acceleration with a past deceleration is a 
possible solution of the Friedmann equation 
by considering the Universe as a 
mixture of a scalar with a matter field and by including a non-equilibrium
pressure term in the energy-momentum tensor. The dark 
energy density decays more slowly with respect to the time 
than the matter energy density does. 
The inclusion of the non-equilibrium pressure leads to a less pronounced decay 
of the matter field with a shorter period of past deceleration.
\end{abstract}


According to recent cosmological observations there exists an evidence
that the Universe is flat (see Sievers et al~\cite{Sie})  and  
expanding with a positive 
acceleration (see Perlmutter et al~\cite{Per} and Riess
et al~\cite{Ri}). The flatness of the Universe is connected with the
total density parameter $\Omega_{\rm tot}$ which 
is the sum of the density parameters related to
vacuum energy $\Omega_{\Lambda}$, cold dark matter  $\Omega_{\rm cdm}$ and 
baryons $\Omega_{b}$, i.e., $\Omega_{\rm tot}=\Omega_{\Lambda}+
\Omega_{\rm cdm}+\Omega_{b}$. The sum of the density parameters of
cold dark matter  and baryons refers to the matter density parameter 
$\Omega_{m}=\Omega_{\rm cdm}+\Omega_{b}$. Most recent values for the 
density parameters from measurements of the anisotropy
of the cosmic microwave background (CMB) are~\cite{Sie}: 
$$\Omega_{\rm tot}=1.01^{+0.09}_{-0.06}\,;
\qquad  \Omega_{m}=0.37\pm0.11\qquad\hbox{and}\qquad \Omega_{b}=0.060\pm0.020.
$$
Measurements of the redshift of the type Ia supernova SN 1997ff indicate 
a present acceleration of the Universe  and - according to Turner and 
Riess~\cite{TR} -
provide a past decelerating phase of the Universe.

Phenomenological cosmologies play an important role in the understanding of the
evolution of the Universe. A remarkable combination of general relativity and  
thermodynamics  allow the description of different regimes 
in cosmological theories. Models that include the recent 
experimental 
evidence of a present  accelerated Universe therefore  deserve an increasing 
attention.

On the other hand,  there exist other regimes where a positive acceleration is
present. Indeed, in  inflationary models  an exponential
expansion with positive acceleration for the early Universe is proposed
 where a hypothetical  particle, the inflaton, is at the core of that 
mechanism.
These ideas play the role of a precursor for the relation between vacuum 
(or dark) 
energy and a  cosmological constant that depends on time.  For recent 
reviews on 
dark energy one is referred to the works by 
Peebles and Ratra~\cite{PR} and Turner~\cite{Tur}.

What we know today is that  dark energy dominates the composition of our
Universe and is the responsible for the positive accelerated expansion. It
interacts weakly with ordinary matter (baryonic  matter represents at
maximum  5\% of the whole mass/energy composition of our Universe).

The point we would like to discuss in this work is that 
a past decelerated with a present accelerated expansion  of the Universe 
can be found as a solution of the Friedmann  equation.   
The Universe is modeled as a mixture of two constituents
namely,  a scalar field  that represents the dark energy and 
a matter field that describes the baryonic matter and the cold dark matter;
in our approach 
the interaction between the different constituents refers only to the one
between the dark energy and the matter via the gravitational field 
which is represented by the cosmic scale factor $a(t)$.
We model this interaction by using the ideas of a
thermodynamic theory, by means of the inclusion of a non-equilibrium 
pressure term  in the energy-momentum tensor 
which represents an irreversible process of energy transfer between the
 matter and the gravitational fields.
 Within the framework of first order Eckart thermodynamic theory the 
non-equilibrium pressure is proportional to the Hubble parameter and 
its proportionality factor is identified with the bulk viscosity, so 
that this model is  referred in the literature as a viscous
cosmological model (see, for example, the works~\cite{Wein}).

Although  the transition from a past deceleration to a 
present acceleration of the Universe
is found as a possible solution of the Friedmann equation 
independently of the presence of the non-equilibrium
pressure term, it is very questionable to model the Universe 
as a perfect 
gas mixture of scalar and matter fields evolving
without dissipative 
effects (for further discussions on this subject one is referred
to the works~\cite{Cal}). 
In this work, it is shown that the dark 
energy density decays more slowly with respect to the time 
than the matter energy density does. These conclusions are valid for
 both cases 
where the non-equilibrium pressure is present or absent.
However, the inclusion of a non-equilibrium pressure leads to a 
less pronounced decay 
of the matter field with a shorter period of past deceleration, since
the non-equilibrium pressure is the responsible for the
energy transfer between the matter and the gravitational 
fields. 

We shall consider a spatially flat, homogeneous and isotropic
Universe described by the Robertson-Walker metric. 
With respect to the four-velocity $U^\mu$ the energy-momentum tensor 
$T^{\mu\nu}$ is decomposed as\footnote{Units have chosen so that $c=1$.}
\be
T^{\mu\nu}=(\rho_X+\rho_m+p_X+p_m+\varpi)U^\mu U^\nu-
(p_X+p_m+\varpi)g^{\mu\nu}.
\ee{2}
where  $ p_X$ is the  pressure of the dark energy, $p_m$ is the 
pressure of the matter field and  $\varpi$ denotes the non-equilibrium 
pressure which is coupled to the irreversible process 
of energy transfer between matter and gravitational fields.

In a comoving frame the conservation law of the energy-momentum tensor
${T^{\mu\nu}}_{;\nu}=0$ leads to the balance equation for the energy density 
\be
\dot\rho_X+\dot\rho_m+3H(\rho_X+\rho_m+p_X+p_m+\varpi)=0,
\ee{3}
where $H=\dot a/a$ is the Hubble parameter and the dot refers to a 
differentiation with respect to the time.

The identification of the dark energy with a scalar field 
allow us to assume  that the balance equation for the
dark energy density is given by
\be
\dot\rho_X+3H(\rho_X+p_X)=0.
\ee{8}
Hence, the balance equation for the dark energy density decouples from that 
for the
energy density of the matter and we have from (\ref{3}) and (\ref{8})
\be
\dot\rho_m+3H(\rho_m+p_m)=-3H\varpi.
\ee{3a}

The equation that connects the evolution of the cosmic scale factor $a(t)$
with the energy densities of the scalar and matter fields  is given by the 
Friedmann
equation
\be
H^2={8\pi G\over 3}(\rho_X+\rho_m),
\ee{5}
where $G$ is the gravitational constant.

In a previous work~\cite{KD1} we have calculated the 
energy-momentum pseudotensor of the gravitational field in a flat 
Robertson-Walker metric and found
\be
T^{00}_G=-{3\over 8\pi G}\left({\dot a\over a}\right)^2.
\ee{19}
If we identify the $T^{00}_G$ with the energy density $\rho_G$ of 
the gravitational field
we can get thanks to the Friedmann equation (\ref{5}) the relationship 
$\rho_G=-(\rho_X+\rho_m)$. 

Now we differentiate  $\rho_G=-(\rho_X+\rho_m)$ with respect to the time 
and  by using the balance equation for the energy density (\ref{3a}) we get 
\be
\dot\rho_G+3H(\rho_G-p_m-p_X)=3H\varpi,
\ee{3b}
which can be interpreted as a balance equation for the energy density of
the gravitational field. Moreover, if we compare equations (\ref{3a}) and 
(\ref{3b}) we
infer that the right-hand side of both equations has a term 
(with opposite sign)  proportional to the
non-equilibrium pressure which is responsible 
to the transfer of energy between the gravitational and matter fields.

It is usual to assume  for the pressure of the  dark energy the equation 
of state $p_X=w_X\rho_X$, with $w_X<-{1/ 3}$ 
(see Peebles and Ratra~\cite{PR}).
Hence, one can obtain by integration of (\ref{8})
\be
{\rho_X\over\rho_X^0}=\left({a_0\over a}\right)^{3(w_X+1)},
\ee{12}
where $\rho_X^0$ and $a_0$ are the values of the dark energy density 
and of the cosmic scale factor at $t=0$ (by adjusting clocks), respectively.
 
In order to determine  the time evolution of the dark energy density 
one has to know the evolution of the cosmic scale 
factor which is determined from the system of equations (\ref{3a}), (\ref{5})
and (\ref{12}). This system of equations   is closed  
by assuming
a relationship between the pressure  and the energy 
density of the matter and a constitutive equation for the non-equilibrium 
pressure $\varpi$. We assume a barotropic equation of state for the pressure
$p_m=w_m\rho_m$, with $0\leq w_m\leq1$ and 
a linear relationship between the non-equilibrium pressure $\varpi$ 
and the Hubble 
parameter $H$ within the framework of Eckart first order thermodynamic theory
\be
\varpi=-3\eta H,\qquad \eta=\alpha(\rho_X+\rho_m).
\ee{11}
In the above equation $\eta$ is the coefficient of bulk viscosity which is 
consider to be
proportional to the energy density of the mixture and $\alpha$ is a constant.

We differentiate the Friedmann equation (\ref{5}) with respect to the time
and get by the use of (\ref{12}) and (\ref{11})
\be
\dot H={3\over 2}\left[{(w_m-w_X)\over1 +\rho_m^0/\rho^0_X}
\left({1\over a}\right)^{3(w_X+1)}+(3\alpha H-w_m-1) H^2\right].
\ee{14}
Above $\rho_m^0$ is the energy density of the matter field at $t=0$ (by 
adjusting clocks). Moreover,  all terms in (\ref{14}) are dimensionless 
quantities defined by
\be
H\equiv {H\over H_0},\quad t\equiv tH_0,\quad a\equiv {a\over a_0},\quad
\alpha\equiv\alpha H_0,\quad\hbox{with}\quad
H_0=\sqrt{{8\pi G \over 3}(\rho_X^0+\rho_m^0)}.
\ee{15}

Equation (\ref{14}) is a second-order differential equation for the cosmic
scale factor $a(t)$ which is a function of four parameters, namely $w_m$,
$w_X$, $\alpha$ and $\rho_m^0/\rho^0_X$. The solution of the differential 
equation  (\ref{14}) is found by specifying values for two initial 
conditions and for the four parameters.

Instead of using the constitutive equation (\ref{11}) for the  
non-equilibrium pressure 
one may 
consider it as a variable within the framework of extended (causal or 
second-order) thermodynamic theory. In this case the evolution equation 
for the non-equilibrium pressure - in a linearized theory - reads
(see, for example~\cite{MR})
\be
\varpi+\tau\dot\varpi=-3\eta H,
\ee{20}
where $\tau$ is a characteristic time. Here we follow 
the works~\cite{Di}
and assume that the characteristic time is given by $\tau=\eta/\rho$.

Hence, we obtain from  the Friedmann equation (\ref{5}) and from the evolution
equation for the non-equilibrium pressure (\ref{20}) the following  
system of differential equations  
\be
\dot H={3\over 2}\left[{(w_m-w_X)\over1 +\rho_m^0/\rho^0_X}
\left({1\over a}\right)^{3(w_X+1)}-(w_m+1) H^2+\varpi\right],
\ee{14a}
\be
\varpi+\alpha\dot\varpi=-3\alpha H^3,
\ee{21}
thanks to (\ref{11})$_2$. Apart from the dimensionless
quantities introduced above,  $\varpi\equiv 8\pi G\varpi/(3H_0^2)$ is 
a dimensionless
non-equilibrium pressure. 

If we specify values for three initial 
conditions and for the four parameters $w_m$,
$w_X$, $\alpha$ and $\rho_m^0/\rho^0_X$ we can obtain from the system 
of equations
(\ref{14a}) and (\ref{21}) the time evolution of  the cosmic 
scale factor $a(t)$ and of the non-equilibrium pressure $\varpi(t)$. 

Once $a(t)$ is determined from
(\ref{14}) or from the system (\ref{14a}) and (\ref{21}) 
one can find the energy densities by using the expressions
\be
\rho_X=\left({1\over a}\right)^{3(w_X+1)},\qquad
\rho_m=\left(1 +{\rho_X^0\over\rho^0_m}\right)H^2
-{\rho_X^0\over\rho^0_m}\left({1\over a}\right)^{3(w_X+1)},
\ee{17}
where $\rho_X\equiv\rho_X/\rho_X^0$ and $\rho_m\equiv\rho_m/\rho_m^0$ are
dimensionless quantities.

The initial conditions we choose at the instant of time $t=0$ (by adjusting 
clocks)  are: $a(0)=1$ for the cosmic
scale factor and  $H(0)=1$ for the Hubble parameter and $\dot\varpi(0)=0$
for the non-equilibrium pressure. 
Since the differential equations (\ref{14}), (\ref{14a}) and (\ref{21}) do
depend on the four parameters $w_m$,
$w_X$, $\alpha$ and $\rho_m^0/\rho^0_X$ there exists much freedom to find 
their solutions.
In order to obtain the graphics in the  Fig. 1 we have chosen:
\begin{itemize}
\item[a)] $w_X=-0.7$ so that  the condition $w_X<-1/3$ holds;
\item[b)] $w_m=0.2$, so that   $0<w_m<1/3$ where the value 
$w_m=0$  refers to dust and $w_m=1/3$ to radiation;
\item[c)] $\rho_m^0/\rho^0_X=2$, i.e., the amount of the matter energy 
density is twice the one of the dark energy density;
\item[d)] $\alpha=0$  when the non-equilibrium pressure is 
absent and  $\alpha=0.05$ (say) when it is present.
\end{itemize}

We have plotted in Fig. 1 the dark energy density $\rho_X$, the matter
energy density $\rho_m$ and the acceleration $\ddot a$ 
as functions of the time $t$ for the three cases: a) by using 
a second-order thermodynamic theory (straight lines); 
b) by using a first-order thermodynamic  theory 
 (dashed lines) and c) by neglecting the non-equilibrium pressure 
(perfect fluid solution). 
We infer from this figure
that for the parameters chosen above the behavior
of the solutions of
the first-order  and second-order thermodynamic theories is practically the 
same. In the three cases there exist periods where the Universe is  
decelerating and accelerating. 
Moreover, the dark energy density decays 
more slowly than the matter energy density does. Even when a small amount of 
dark energy density - with respect to the matter energy  density - 
is taken into 
account, these fields evolve in such a manner that for large times the 
amount of dark energy density is very large with respect to the energy density 
of the matter field.  The inclusion of the non-equilibrium pressure 
lead to: i) a slower decay of the matter field with respect to the time, 
which is a consequence
of the irreversible process of energy transfer between the matter 
and gravitational fields; ii) a smaller interval 
of past deceleration and iii) a more rapid decay of the dark energy density. 

Other conclusions can be obtained by changing the values of the parameters.
In the case where there is no dark energy density $\rho_X=0$ there is
no period of acceleration, i.e., only a period of deceleration is possible. By
changing the interval $w_m$ to $1/3<w_m<2/3$ - which corresponds to 
the interval
between radiation $w_m=1/3$ and non-relativistic matter $w_m=2/3$ 
- the period of past deceleration 
increases. This last result is also found when $w_X$ increases, i.e., for
$w_X>-0.7$. By increasing  the value of the dimensionless constant 
$\alpha$ the effect of the non-equilibrium pressure is more pronounced and
the period of past deceleration decreases.
   
\begin{figure}
\begin{center}
\includegraphics[width=8.5cm]{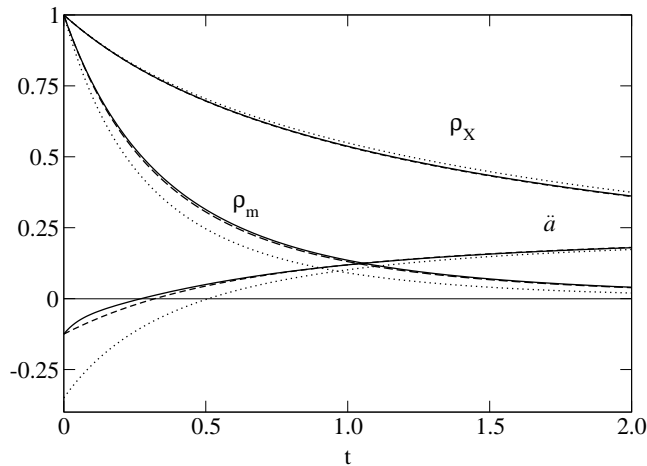}
\caption{Dark energy density  $\rho_X$, matter energy density $\rho_m$
and acceleration $\ddot a$ vs time $t$ for perfect fluid
(dotted line), first-order 
 (dashed line)
and second-order (straight line) thermodynamic theories.}
\end{center}
\end{figure}

\end{document}